\documentclass[conference]{IEEEtran}
\IEEEoverridecommandlockouts

\usepackage[nolist]{acronym}
\usepackage{url}
\usepackage{cite}
\usepackage{booktabs}
\usepackage{amsmath,amssymb,amsfonts}
\usepackage{hyperref}
\usepackage{algorithmic}
\usepackage{graphicx}
\usepackage{textcomp}
\usepackage{xcolor}

\def\BibTeX{{\rm B\kern-.05em{\sc i\kern-.025em b}\kern-.08em
    T\kern-.1667em\lower.7ex\hbox{E}\kern-.125emX}}
\begin{document}

\title{An Octave-based Multi-Resolution C$Q$T Architecture for Diffusion-based Audio Generation\\

\thanks{Both authors contributed equally to this work. It was funded by Volkswagen Foundation (Volkswagen Stiftung) Germany, under Grant no. 96 881.
}
}

\author{\IEEEauthorblockN{
Maurício do V. M. da Costa}
\IEEEauthorblockA{\textit{MTDML, IMM} \\
\textit{University of Osnabrück}\\
Osnabrück, Germany \\
madovalemade@uni-osnabrueck.de}
\and
\IEEEauthorblockN{Eloi Moliner}
\IEEEauthorblockA{
\textit{Acoustics Lab, DICE} \\
\textit{Aalto University} \\
Espoo, Finland \\
eloi.moliner@aalto.fi}
}

\begin{acronym}
\acro{stft}[STFT]{short-time Fourier transform}
\acro{istft}[iSTFT]{inverse short-time Fourier transform}
\acro{dnn}[DNN]{deep neural network}
\acro{pesq}[PESQ]{Perceptual Evaluation of Speech Quality}
\acro{polqa}[POLQA]{perceptual objectve listening quality analysis}
\acro{wpe}[WPE]{weighted prediction error}
\acro{psd}[PSD]{power spectral density}
\acro{rir}[RIR]{room impulse response}
\acro{hrir}[HRIR]{head-related impulse response}
\acro{brir}[BRIR]{binaural room impulse response}
\acro{hrtf}[HRTF]{head-related transfer function}
\acro{snr}[SNR]{signal-to-noise ratio}
\acro{lstm}[LSTM]{long short-term memory}
\acro{polqa}[POLQA]{Perceptual Objective Listening Quality Analysis}
\acro{sdr}[SDR]{signal-to-distortion ratio}
\acro{cnns}[CNNs]{Convolutional Neural Networks}
\acro{vae}[VAE]{variational auto-encoder}
\acro{gan}[GAN]{generative adversarial network}
\acro{tf}[T-F]{time-frequency}
\acro{sde}[SDE]{stochastic differential equation}
\acro{ode}[ODE]{ordinary differential equation}
\acro{drr}[DRR]{direct to reverberant ratio}
\acro{lsd}[LSD]{log spectral distance}
\acro{sisdr}[SI-SDR]{scale-invariant signal to distortion ratio}
\acro{mos}[MOS]{mean opinion score}
\acro{map}[MAP]{maximum a posteriori}
\acro{sde}[SDE]{stochastic differential equation}
\acro{ode}[ODE]{ordinary differential equation}
\acro{dps}[DPS]{diffusion posterior sampling}
\acro{tfrs}[TFRs]{Time-Frequency Representations}
\acro{cqt}[C$Q$T]{Constant-$Q$ Transform}
\acro{ldms}[LDMs]{Latent Diffusion Models}
\end{acronym}

\maketitle

\begin{abstract}
This paper introduces MR-CQTdiff, a novel neural-network architecture for diffusion-based audio generation that leverages a multi-resolution Constant-$Q$ Transform (C$Q$T). The proposed architecture employs an efficient, invertible CQT framework that adjusts the time-frequency resolution on an octave-by-octave basis. This design addresses the issue of low temporal resolution at lower frequencies, enabling more flexible and expressive audio generation. We conduct an evaluation using the Fréchet Audio Distance (FAD) metric across various architectures and two datasets. Experimental results demonstrate that MR-CQTdiff achieves state-of-the-art audio quality, outperforming competing architectures.
\end{abstract}


\section{Introduction}

\ac{tfrs} are fundamental tools in a wide range of audio processing applications, often employed in state-of-the-art deep learning systems for audio analysis \cite{rafii2022constant, riou2023pesto }, synthesis~\cite{marafioti2019adversarial, donahueadversarial, engelgansynth, huangtimbretron}
, and enhancement~\cite{richter2023speech, moliner2023solving}.
\ac{tfrs} can be computed with varying resolutions. 
For example, generating a spectrogram using the \ac{stft} involves selecting an analysis window whose length determines the trade-off between time and frequency resolution.
The longer the analysis window, the higher the frequency resolution—at the cost of lower time resolution—and vice versa. Therefore, selecting an appropriate resolution for the task and signal characteristics is essential.

Spectrograms based on the \ac{stft} use linear time and frequency grids. For audio with strong harmonic content, such as music, harmonic components of musical notes are linearly spaced in frequency, and their spacing varies across pitches.
An alternative TFR, the \ac{cqt}, adopts a logarithmic frequency scale, yielding pitch-invariant harmonic spacing. This makes the C$Q$T particularly suitable for \ac{cnns}, as it aligns well with the translation-equivariant nature of convolutional kernels when processing harmonic signals.
However, the \ac{cqt} exhibits low time resolution at lower frequencies, which limits its ability to capture transient events. This leads to excessive temporal smearing of percussive sounds and low-pitch notes, resulting in degraded feature quality for learning tasks.

This paper focuses on diffusion-based generative models \cite{ho2020denoising, song2020score, karras2022elucidating}, a class of deep generative methods that have demonstrated strong performance across various tasks,
including speech synthesis \cite{popov2021grad, jeong2021diff},
audio restoration \cite{lemercier2025diffusion}, and automatic music generation \cite{liu2023audioldm,
evans2025stable}. 
Although diffusion models are architecture-agnostic, both architectural design and the choice of data representation significantly affect performance. Representations that emphasize task-relevant structure introduce inductive biases that facilitate training and improve sample quality~\cite{moliner2024diffusion}.

Early approaches to diffusion-based audio generation operated directly on the waveform domain, targeting applications such as speech \cite{chenwavegrad}, sound effects \cite{pascual2023full}, and drum synthesis \cite{rouard2021crash}. However, modeling raw waveforms remains challenging due to their high dimensionality and limited structure, making pattern learning inefficient and poorly scalable. As an alternative, several works proposed operating in the mel-spectrogram domain, which provides a lower-dimensional and perceptually meaningful representation of audio \cite{koizumi2022specgrad, wu2024music, popov2021grad}. While this reduces the complexity of modeling, it introduces an additional burden: the need for a separately trained neural vocoder to reconstruct audio, which can become a bottleneck in quality and flexibility.

More recently, following their success in image generation \cite{rombach2022high}, latent diffusion models (LDMs) have emerged as a dominant approach for audio generation \cite{evans2025stable, liu2023audioldm, schneider2023mo, nistal2024diff}. These models decompose the generative process into two stages: first, an autoencoder is trained to compress audio into a lower-dimensional latent space; then, a diffusion model is trained to model the distribution of these latent representations. This strategy concentrates domain-specific design efforts on the autoencoder, allowing the diffusion model to adopt general-purpose, highly scalable architectures, such as transformers \cite{peebles2023scalable}, due to the reduced structure and dimensionality of the latent space. While this approach enables efficient training and the modeling of complex audio distributions, it also presents notable challenges. The quality of the final output is inherently limited by the reconstruction error of the autoencoder, and the separation of training into two independent stages introduces additional complexity~\cite{dieleman2025latents}. Moreover, operating in a latent domain often complicates conditional generation tasks, such as solving inverse problems, where the relationship between observations and latent variables is indirect or unknown \cite{daras2024survey}.

An alternative paradigm seeks to overcome the curse of dimensionality by leveraging invertible time–frequency representations. These representations reveal inherent structure in audio signals, exhibit sparsity, and provide a potentially more tractable domain for modeling complex data distributions.
One line of work explores the use of the \ac{stft}, either by defining the diffusion process directly in the STFT domain \cite{richter2023speech, kong2025a2sb}, or by exploiting the differentiability of the inverse STFT to operate in the waveform domain while processing features in the time–frequency domain \cite{moliner2023solving, lemercier2025unsupervised}. The latter approach retains the simplicity of waveform-based modeling, while benefiting from the inductive biases introduced by time–frequency representations.
Building on this idea, Moliner et al.~\cite{moliner2023solving, moliner2024diffusion} proposed using the \ac{cqt} for music restoration tasks. Their method relies on an invertible and differentiable implementation of the \ac{cqt}, combined with a U-Net architecture that exploits temporal redundancies in the transform for efficient and scalable processing \cite{moliner2024diffusion}.

In this work, we introduce the multi-resolution CQTdiff (MR-CQTdiff), a modification to the CQTdiff+ model proposed in \cite{moliner2024diffusion}. Our architecture leverages a C$Q$T filter bank, i.e. multiple parallel C$Q$Ts with different resolutions covering complementary frequency ranges, aiming to balance time-frequency resolution by progressively decreasing the ratio between time and frequency resolutions on an octave basis across the audible spectrum. In doing so, it avoids excessively low time resolution at low frequencies while maintaining a relatively high frequency resolution at high frequencies, which a single C$Q$T cannot achieve. While still supporting a U-Net-based architecture, the C$Q$T filter bank better captures transient and low-frequency content, enabling higher-quality generation of musical material overall.

The remainder of the paper is organized as follows: Section~\ref{sec:diffusion} introduces the diffusion model framework and training strategy; Section~\ref{sec:MR-CQT} details the MR-CQTdiff architecture; Section~\ref{sec:Evaluation} presents the experiments conducted and results achieved; and Section~\ref{sec:Conclusions and Future Work} concludes with a discussion of findings and future research directions.

\section{Diffusion Models for Audio Generation}\label{sec:diffusion}

Diffusion models are a class of generative models that learn to synthesize data by reversing a gradual noising process. In the forward process, data samples 
$\boldsymbol{x}_0$
are progressively corrupted by adding Gaussian noise, resulting in a sequence of increasingly noisy versions of the data $\boldsymbol{x}_\tau$. This transformation defines a diffusion process parameterized by a continuous time variable $\tau$~\cite{song2020score}.\footnote{Note that the time variable $\tau$ is used to describe the diffusion process, which is independent of the time domain $t$ of the audio signal.}
During training, a neural network is optimized to approximate the
score function $s_\theta(\boldsymbol{x}, \tau) \approx
\nabla_{\boldsymbol{x}} \log p_\tau(\boldsymbol{x}_\tau)$, which defines 
a vector field that indicates the direction towards regions of higher probability at time $\tau$.
 Once trained, the score model can be used to guide a reverse diffusion process, 
which transforms samples drawn from of Gaussian distribution 
$\boldsymbol{x}_T \sim \mathcal{N}(\boldsymbol{0}, \sigma^2_{\mathrm{max}}\boldsymbol{I})$ into samples from the 
training data distribution  $\boldsymbol{x}_0 \sim p_\mathrm{data}$.

The score model $s_\theta(\mathbf{x}_\tau, \tau)$ is typically trained using the denoising score matching objective \cite{vincent2011connection}:
\begin{equation} \label{eq:objective}
    \mathbb{E}_{\boldsymbol{x}_0, \boldsymbol{\varepsilon} \sim \mathcal{N}(\mathbf{0}, \mathbf{I})} \left[ \lambda(\tau) \left\| s_{\boldsymbol\theta}(
    \boldsymbol{x}_0 + \sigma(\tau)\boldsymbol{\varepsilon}, \tau)  
    -
     \frac{
     \boldsymbol{x}_0 -\boldsymbol{x}_\tau
     }{\sigma^2(\tau)} \right\|^2_2 \right],
\end{equation}
where $\lambda(\tau)$ is a time-dependent weighting parameter, and $\sigma(\tau)$ defines the noise level at time $\tau$.

Following \cite{moliner2023solving, moliner2024diffusion}, we adopt several design choices proposed by Karras et al. \cite{karras2022elucidating}, such as defining $\sigma(\tau)=\tau$.
To ensure stability and scale consistency throughout training, Karras et al.~\cite{karras2022elucidating} also propose the following parameterization of the score model: 
\begin{equation}
    s_{\boldsymbol{\theta}}(\boldsymbol{x}_{\tau}, \tau) =
    \frac{
    (c_{\text{skip}}(\tau)-1)x_{\tau} + c_{\text{out}}(\tau) F_{\boldsymbol{\theta}}(c_{\text{in}}(\tau)\boldsymbol{x}_{\tau}, \tau)
    }
    {\sigma^2(\tau)}
    ,
\end{equation}
where $F_{\boldsymbol{\theta}}$ is the core neural network, 
and the weighting parameters $c_{\text{skip}}$, $c_{\text{out}}$ and $c_{\text{in}}$ are chosen to maintain close-to-unit variance in the input and output of $F_{\boldsymbol{\theta}}$, 
, which is known to improve neural network training stability.

The contributions of this paper focus on the architectural design of the core network $F_{\boldsymbol{\theta}}$. While our experiments adopt the parameterization introduced by Karras et al.~\cite{karras2022elucidating}, we believe that architectural choices are largely orthogonal to the specific diffusion parameterization. Therefore, our findings should generalize to alternative formulations, such as DDPM\cite{ho2020denoising} or Flow Matching~\cite{lipmanflow}.

\subsection{Inference}\label{sec:inference}

Also following \cite{karras2022elucidating}, we employ the following \ac{ode} to traverse the generative (reverse) process:
\begin{equation}
d\boldsymbol{x} = -\tau \, s_\theta(\boldsymbol{x}_\tau, \tau) \, d\tau,
\end{equation}
where $d\tau$ is an infinitesimal negative time step.

At inference time, the continuous time variable $\tau$ is discretized into a sequence of $T$ steps using a noise schedule defined as
\begin{equation}\label{schedule}
\tau_i = \left( \sigma_{\mathrm{max}}^{1/\rho} + \frac{i}{T-1} \left( \sigma_{\mathrm{min}}^{1/\rho} - \sigma_{\mathrm{max}}^{1/\rho} \right) \right)^\rho, \quad i = 0, \dots, T-1,
\end{equation}
where $\sigma_{\mathrm{max}}$ and $\sigma_{\mathrm{min}}$ denote the maximum and minimum noise levels, respectively, and $\rho$ controls the nonlinearity of the spacing between time steps. Unless otherwise stated, we use $\sigma_{\mathrm{max}} = 8$, $\sigma_{\mathrm{min}} = 10^{-5}$, $\rho = 10$, and $T = 51$ steps.

To numerically integrate the reverse ODE, we use the second-order Heun's method (also known as improved Euler), as proposed in \cite{karras2022elucidating}. This solver provides a good trade-off between sample quality and computational efficiency.

\section{Multi-Resolution C$Q$T Architecture}\label{sec:MR-CQT}

Our architecture builds upon the CQT-Diff+ algorithm, proposed in~\cite{moliner2024diffusion}, which operates in the time-frequency domain using a \ac{cqt}. The transform implementation is a differentiable version of the C$Q$T introduced by Velasco et al.~\cite{velasco2011constructing} and Holighaus et al.~\cite{holighaus2012framework}, which is computationally efficient by leveraging FFT-based band-pass filters. This implementation is also invertible, enabling perfect reconstruction up to numerical error.

Formally, the system can be described as
\begin{equation}
    F_{\boldsymbol\theta} = \mathrm{IC}Q\mathrm{T} \circ U_{\boldsymbol\theta} \circ \mathrm{C}Q\mathrm{T},
\end{equation}
where $\circ$ denotes the function composition operation, $U_{\theta}$ represents the neural network with trainable weights $\theta$, and C$Q$T and IC$Q$T are the constant-$Q$-transform operator and its inverse, respectively. Note that the IC$Q$T must be \textit{differentiable} to allow backpropagation, but there is no need to make it \textit{trainable}.

The center frequencies $f_k$ of the $K$ filters $g_k$ are logarithmically distributed within the frequency range of interest and can be calculated by
\begin{equation}
    f_k = f_{\mathrm{min}}2^{\frac{k-1}{b}},~\mathrm{for}~k = 1,2,3,\dots,K,
\end{equation}
where $b$ denotes the number of bins per octave and $f_{\mathrm{min}}$ is the lowest center frequency. The maximum frequency can be arbitrarily chosen and is typically set close to the Nyquist limit $f_k = f_\mathrm{s}/2$. In our implementation, different C$Q$Ts cover complementary frequency ranges, thus requiring different minimum and maximum frequencies.

Given the strong presence of harmonic content in musical signals, the constant-$Q$ transform (C$Q$T) provides a key advantage: its logarithmic frequency scale promotes pitch-equivariant symmetry. This property makes the C$Q$T particularly well-suited for convolutional architectures, outperforming general STFT-based spectrograms in harmonic contexts. Due to the transform’s design (where filters become progressively narrower at lower frequencies), the corresponding impulse responses in the time domain must be proportionally longer, as dictated by the uncertainty principle.\footnote{The uncertainty principle, or Heisenberg's principle, states that the product of a signal’s time support and the frequency support of its transform is lower bounded.} An additional consequence is that a C$Q$T with a regular time-frequency grid inherently exhibits redundancy in the time domain, which increases toward lower frequencies.

An effective strategy to reduce excessive redundancy, used in CQT-Diff+, is to adopt octave-based regular grids, where the number of time frames is halved with each lower octave. This approach integrates naturally with the U-Net architecture, which represents data at multiple resolutions through progressive downsampling. In CQT-Diff+, the first U-Net level processes the highest octave (with the finest time resolution), and each subsequent level receives a concatenation of the next lower octave with a downsampled version of the higher octaves. This ensures that all inputs are aligned in resolution. Further implementation details are provided in~\cite{moliner2024diffusion}.

Despite its efficiency, this system inherits a fundamental limitation of the C$Q$T: poor time resolution at low frequencies. This manifests as increasing temporal smearing toward the lower end of the spectrum, which compromises the representation of transient information, such as fast pitch fluctuations. One way to improve time resolution at low frequencies is to reduce the number of bins per octave ($b$), resulting in shorter filter impulse responses and broader frequency bandwidths. However, this comes at the cost of frequency resolution at higher frequencies, potentially impairing the network’s ability to distinguish fine harmonic structures in that range.

\subsection{Architecture Design}

\begin{figure*}[t]
    \centering
    \includegraphics[width=0.75\linewidth]{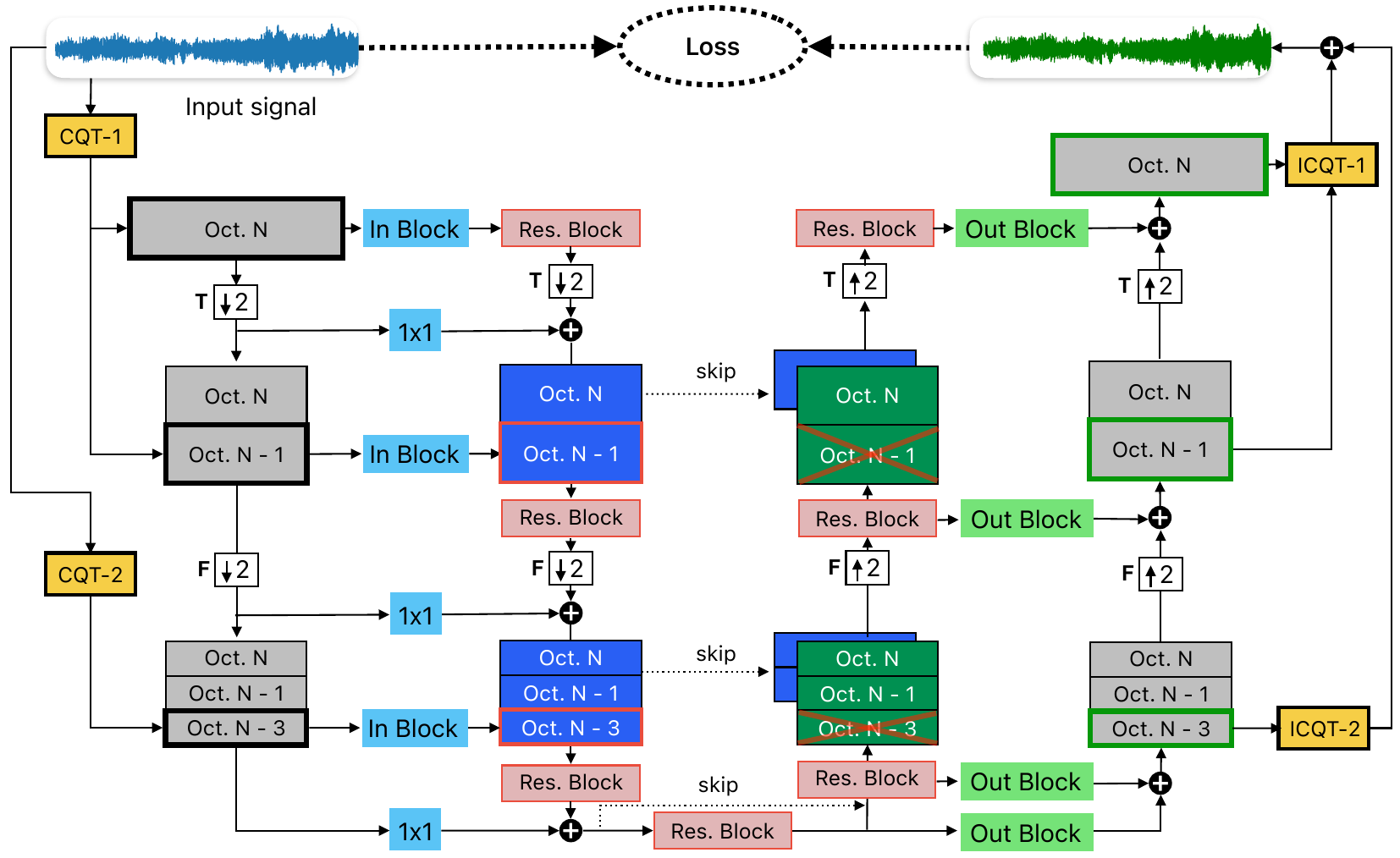}
    \caption{Main structure of the MR-C$Q$Tdiff architecture, illustrating a simplified example with three hypothetical octaves processed by two C$Q$Ts: CQT-1 with resolution $b$ bins per octave, and CQT-2 with $b/2$ bins per octave.}
    \label{fig:Model}
\end{figure*}

The MR-CQTdiff architecture follows a U-Net structure, as illustrated in Figure~\ref{fig:Model}, with concatenative skip connections linking encoder and decoder layers at corresponding resolutions. Anti-aliasing filters are applied during both downsampling and upsampling stages. Each resolution level (including the bottleneck) contains a residual block, referred to as ``Res. Block'', which serves as the core computational unit. In the encoder, the input is divided into octave-specific segments and processed independently using ``In. Blocks''. These features are concatenated along the frequency axis with the corresponding U-Net latents and augmented with residual connections from resized input features to maintain information flow.

The decoder mirrors this dual-path structure: a main path containing ``Res. Blocks'' with progressive upsampling, and an auxiliary ``outer'' path that enhances gradient flow. At each decoder level, lower-octave features are discarded from the main path and passed through ``Out. Blocks'' to the outer path. These features are eventually routed to the IC$Q$T module. 
This two-path setup ensures both efficient feature reuse and improved training stability.

Our solution addresses the aforementioned time-resolution problem by computing multiple C$Q$Ts that cover complementary frequency ranges. The proposed architecture follows the overall structure of CQT-Diff+, replacing certain time-domain downsampling operations with frequency-domain ones when transitioning between resolutions. The main diagram in Figure~\ref{fig:Model} illustrates a simplified example of the architecture for $N = 3$ octaves. In this case, two C$Q$Ts are used: C$Q$T-1 is computed with $b$ bins per octave, while C$Q$T-2 uses $b/2$ bins per octave, trading frequency resolution for improved time resolution at the lowest octave (Oct. $N$–3). Since Oct. $N$ (the highest) and Oct. $N$–1 share the same number of frequency bins $b$, a $2\times$ downsampling in time is applied to match their dimensions for concatenation. At the transition to the next U-Net level---the crossover point between the different C$Q$Ts---the time dimensions of the features are already aligned, so frequency downsampling is applied at that point.

All building blocks in the architecture are conditioned on a noise-level embedding ($\sigma$-emb), constructed using Random Fourier Features \cite{tancik2020fourier}   followed by a three-layer MLP. Conditioning is applied via feature-wise linear modulation without shifts. The ``In. Block'' expands the input (real and imaginary parts) from two channels to the desired latent size using a $1\times1$ convolution, followed by Group Normalization (shift-free), Gaussian-error-linear-unit (``GELU'') activation, and a linear layer. The ``Out. Block'' mirrors this structure, applying $1\times1$ convolution at the end to reduce the latent size back to two channels. Each "Res. Block" contains shift-free Group Normalization, GELU activation, and convolutions in time and frequency, with exponentially increasing dilation along the frequency axis to achieve pitch-equivariant receptive fields. 

\subsection{Hyperparameter Specification}

The system operates with audio signals sampled at 44.1\,kHz, which are transformed by three C$Q$Ts  with resolutions $b = \{8, 16, 32\}$\,bins/oct., respectively covering the frequency ranges $43.1-344.5$\,Hz, $344.5-5512.5$\,Hz, and $5512.5-22050$\,Hz, which together span $N_\mathrm{oct}=9$\,octaves. As a means of comparison, in the original study~\cite{moliner2024diffusion}, the CQTdiff+ algorithm used a C$Q$T with $b=64$, which yields $8\times$ lower time resolution at the first octaves. Table~\ref{tab:Model_config} details the frequency range covered by each octave, its number of bins per octave, the level at the U-Net at which that octave is fed, and the resampling type (time- or frequency-wise).

The U-Net depth corresponds to the number of octaves, with feature sizes increasing from 32 in the shallowest layers to 256 at the bottleneck. Each ``Res. Block'' contains between two and five stacked dilated convolutions, with fewer dilations in shallower layers due to the smaller number of frequency bins and reduced need for a large receptive field.
Importantly, these architectural modifications only marginally affect the overall model size, which remains around 40 million parameters.\footnote{Our implementation is available online on \url{https://github.com/eloimoliner/MR-CQTdiff}}

\begin{table}[t]
    \caption{Hyperparameters of the MR-C$Q$T structure.}
    \centering
    \begin{tabular}{c|cccc}
        \toprule
        \textbf{Octave} & \textbf{Frequency Range (Hz)} & $b$ & \textbf{U-Net Level} & \textbf{Resampling} \\
        \midrule
        9 & 11025 -- 22050   & 32 & 1 & Time \\
        8 & 5512.5 -- 11025  & 32 & 2 & Freq. \\
        \midrule
        7 & 2756.3 -- 5512.5 & 16 & 3 & Time \\
        6 & 1378.1 -- 2756.3 & 16 & 4 & Time \\
        5 & 689.1  -- 1378.1 & 16 & 5 & Time \\
        4 & 344.5  -- 689.1  & 16 & 6 & Freq. \\
        \midrule
        3 & 172.3  -- 344.5  & 8  & 7 & Time \\
        2 & 86.1   -- 172.3  & 8  & 8 & Time \\
        1 & 43.1   -- 86.1   & 8  & 9 & -- \\
        \bottomrule
    \end{tabular}
    \label{tab:Model_config}
\end{table}

\section{Evaluation}\label{sec:Evaluation}

\subsection{Training Datasets}

We trained all models on two datasets: FMA-Large\cite{deff2017fma}, a diverse collection of 106,574 30-second music tracks across various genres, and OpenSinger\cite{huang2021multi}, a dataset of professionally recorded solo vocal performances. These datasets present distinct challenges: FMA-Large covers a broad range of musical styles and production qualities, while OpenSinger focuses on clean vocal recordings across different singers and pitches, making them complementary benchmarks for evaluating generative performance.\footnote{We opted not to run experiments with the MAESTRO dataset due to the presence of poor-quality, noisy recordings, especially in early recordings. Preliminary tests resulted in noisy samples across all models, significantly influencing in the quality assessment of the generated samples.}

\subsection{Baselines}

We trained four models in the waveform domain using 6-second audio segments and the same diffusion parameterization, formally described in Section~\ref{sec:diffusion}. The only difference between models lies in the architectural choices:
\begin{itemize}
\item \textit{UNet-1D}: A 1-dimensional U-Net composed of temporal convolutions, similar to architectures used in waveform-domain diffusion  \cite{rouard2021crash, pascual2023full, chenwavegrad}.
\item \textit{NCSN++}: A 2-dimensional U-Net operating on STFT representations, originally introduced in \cite{song2020score} and later adapted for speech enhancement in \cite{richter2023speech}. This architecture has also been applied to speech and singing voice modeling \cite{lemercier2025unsupervised}, following the same differentiable composed design. It uses 2D convolutions over time and frequency axes.
\item \textit{CQTdiff+}: The baseline model introduced in \cite{moliner2023solving}, which uses a differentiable and invertible C$Q$T representation combined with a U-Net architecture. Our proposed model is built upon this baseline.
\item \textit{MR-CQTdiff} (ours): The proposed model, which extends CQTdiff+ by introducing a multi-resolution C$Q$T filter bank.
\end{itemize}
All these architectures are configured to share a similar parameter count of around $40$ million parameters.

To provide additional comparison, we also evaluated a latent diffusion model (\textit{LDM}) using the publicly available autoencoder from Stable Audio Open \cite{evans2025stable}. This model employs a Transformer-based architecture similar to that used in the original work, and has approximately 67 million parameters. To ensure a fairer comparison, we adapted our diffusion parameterization to the latent domain, applying the same training framework as used for the waveform-domain models. Necessary adjustments, such as modifying the noise schedule, training loss weighting, and preconditioning, were made to account for the properties of the latent space.

All models were trained for 500{,}000 iterations using the Adam optimizer with a learning rate of $1 \times 10^{-4}$ and a batch size of 4, on a single NVIDIA A100 GPU. Model checkpoints were saved and evaluated every 100{,}000 iterations.
During training, we maintained an exponential moving average (EMA) of the model weights with a decay rate of 0.9999, which was used for inference. 
Total training times (500{,}000 iterations) varied depending on the model: approximately 46 hours for \textit{UNet-1D}, 80 hours for the \textit{NCSN++} model, 125 hours for \textit{CQTDiff+}, 120 hours for \textit{MRCQT-Diff}, and 25 hours for \textit{LDM}.

\subsection{Experiments: Unconditional Generation}

We evaluated the performance of the five models on unconditional audio generation using two datasets: \textit{OpenSinger} and \textit{FMA}. Generation was performed using the same sampler and sampling parameters described in Section~\ref{sec:inference}, with the exception of the \textit{LDM}, for which we used $\sigma_\mathrm{max} = 100$ and $\sigma_\mathrm{min} = 10^{-4}$. For each model, we unconditionally generated 512 audio samples of 6 seconds every 100{,}000 training iterations.

To assess generation quality, we used the \textit{Fréchet Audio Distance (FAD)}, computed from CLAP embeddings~\cite{elizalde2023clap} using the official FADtk implementation~\cite{fadtk}.
FAD quantifies the distance between the distributions of embeddings extracted from real and generated audio.
 In order to analyze the contribution of each test sample to the FAD, and thus potentially detect outliers, we compute individual FAD scores for each song, as proposed in~\cite{fadtk}. CLAP embeddings were selected due to their strong correlation with perceptual audio and musical quality~\cite{fadtk}.

As the reference set for computing FAD, we used the held-out test partitions of \textit{FMA} and \textit{OpenSinger}. The \textit{FMA} test set comprises approximately 17 hours of music, while the \textit{OpenSinger} test set contains around 2.5 hours of audio. Since the reference sets are entirely disjoint from the training data, this evaluation setup allows us to measure the models' generalization ability.


\subsection{Results}

\begin{figure}[t]
    \centering
    \includegraphics[width=0.98\linewidth]{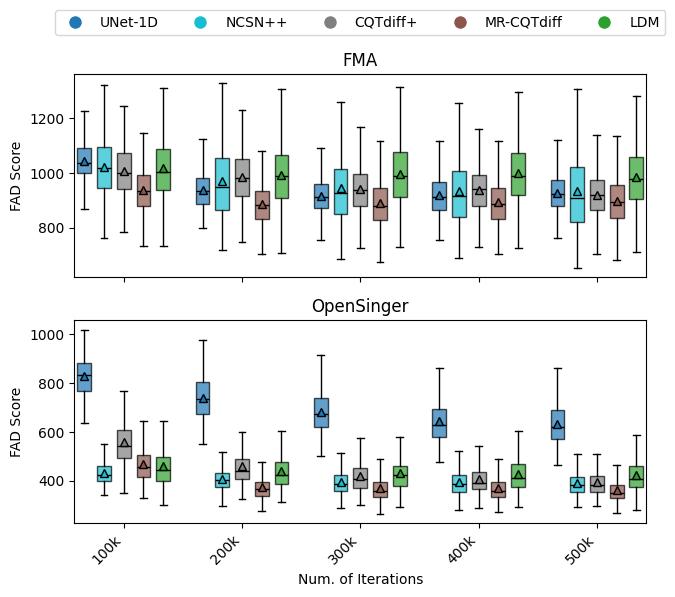}
      \caption{
        Boxplot of per-example FAD scores computed with CLAP embeddings, across different models and training iterations, for the \textit{FMA} and the \textit{OpenSinger} datasets. Lower values indicate closer alignment between generated and real audio distributions.}
    \label{fig:results}
\end{figure}

Figure~\ref{fig:results} illustrates the distributions of FAD scores for all models at various training iterations, separately for the \textit{FMA} (top) and \textit{OpenSinger} (bottom) datasets.
For the \textit{OpenSinger} dataset, the proposed architecture clearly outperforms the others, achieving near-stable performance by around 200,000 iterations. This suggests that the model effectively captures transient details, which are especially important for singing voice, where rapid pitch variations and prominent non-harmonic sounds (e.g., consonants) occur frequently.

A similar trend is observed in the results obtained using the \textit{FMA} dataset, although the score distributions are wider, reflecting the greater diversity and complexity of general music data. Despite this challenge, the proposed MR-CQTdiff consistently attains the lowest median FAD scores and stabilizes its performance by 200{,}000 iterations, whereas other models require more iterations to converge. In this setting, the latent diffusion model (LDM) performs the worst, possibly due to artifacts introduced by the autoencoder reconstruction.

Qualitative audio examples are available on the companion website\footnote{
\href{https://eloimoliner.github.io/MR-CQTdiff/}{eloimoliner.github.io/MR-CQTdiff/}
}. Although samples generated from the \textit{OpenSinger} model, especially MR-CQTdiff, exhibit high quality, examples from the \textit{FMA} dataset often sound less pleasing across all models. This outcome is expected given the higher variability and complexity of the \textit{FMA} dataset and the unconditional generation task. We hypothesize that more perceptually satisfying results could be achieved through guided or conditional generation approaches.

\section{Conclusions and Future Work}\label{sec:Conclusions and Future Work}

In this work, we introduced the MR-CQTdiff: a novel architecture for diffusion-based audio generation that leverages a multi-resolution constant-$Q$ transform. The proposed architecture aims to overcome the time-frequency resolution trade-offs inherent to standard C$Q$T-based representations. 

We evaluated the proposed architecture using the Fréchet Audio Distance (FAD) computed with CLAP embeddings on the \textit{FMA} and the \textit{OpenSinger} datasets. The results obtained show that our model consistently achieves competitive or superior FAD scores compared to four strong baselines: UNet-1D, NCSN++, CQTdiff+, and LDM. In particular, MR-CQTdiff outperforms all other models across most training stages, especially on the OpenSinger dataset, demonstrating improved handling of harmonically rich and transient vocal content.

It is important to note that our experiments were conducted on relatively small-scale models. While latent diffusion models (LDMs) offer clear advantages in scalability and training efficiency, they still underperform MR-CQTdiff at comparable model sizes and training iterations. This highlights the potential of time-frequency-based approaches in high-fidelity audio generation, despite imposing scalability challenges. Future work could explore making MR-CQTdiff more efficient and scaling it to larger architectures and datasets, besides conducting a more comprehensive subjective quality assessment through controlled listening tests.

Finally, we emphasize that the focus of this study was on model performance rather than computational efficiency. Although LDMs remain more efficient in terms of resource usage, our results show that MR-CQTdiff is a compelling alternative when quality is the primary concern.

\bibliographystyle{IEEEtran}
\bibliography{references}
\end{document}